# Tailoring composite skyrmionic spin textures in an above-room-temperature ferromagnet $Fe_{3-x}GaTe_2$

Songyang Li[1,2,+], Jianfeng Guo[1,2,3,+], Zizhao Gong[3,+], Guojing Hu[3,+], Shuo Mi[1,2], Chang Li[1,2], Yanyan Geng[1,2], Manyu Wang[1,2], Shumin Meng[1,2], Shiyu Zhu[3], Fei Pang[1,2], Wei Ji[1,2], Rui Xu[1,2,*], Haitao Yang[3,*], and Zhihai Cheng[1,2,*]

[1]*Key Laboratory of Quantum State Construction and Manipulation (Ministry of Education), School of Physics, Renmin University of China, Beijing 100872, China*

[2]*Beijing Key Laboratory of Optoelectronic Functional Materials & Micro-nano Devices, School of Physics, Renmin University of China, Beijing 100872, China*

[3]*Institute of Physics, Chinese Academy of Sciences, Beijing 100190, China*

**Abstract:** Realizing room-temperature tunable skyrmionic objects in van der Waals ferromagnet offers unparalleled prospects for future spintronics. Here, we report an experimental investigation on the emergence and evolution of skyrmionic spin textures in the non-stoichiometric $Fe_{3-x}GaTe_2$ using magnetic force microscopy. The iron-deficiency-specific magnetic states of stripe, striped skyrmionium and striped skyrmion sack are observed. Through zero-field-cooling and field-cooling measurements, we observed distinct topological transitions and trivial transitions (distinguished by changes in topological charge) emerging during the stepwise evolution of topological spin textures, which enabled us to develop an evolution pathway model. Leveraging this model, the room-temperature stable composite topological spin textures of skyrmionium, skyrmion bag and sack states are further controllably realized via the exclusive topological-transition path (regulated by magnetic field and DMI intensity). Our work provides valuable insights into the room-temperature realization of topological spin textures in $Fe_{3-x}GaTe_2$, and inspires further exploration of their potential applications in heterostructure spintronics.

+ These authors contributed equally to this work.

* To whom correspondence should be addressed:

zhihaicheng@ruc.edu.cn; htyang@iphy.ac.cn; ruixu@ruc.edu.cn

**Introduction**

Topological spin textures including skyrmions, merons, and hopfions have recently garnered significant research attention across diverse condensed matter systems exhibiting broken inversion symmetry, particularly in magnetic systems [1-5], ferroelectric materials [6], and chiral liquid crystals [7]. These topological entities and their derivative configurations (appearing as isolated particles or ordered lattices [8,9]) serve as platforms for emergent phenomena, driving fundamental challenges in understanding their nucleation dynamics and non-equilibrium responses to external fields. Of particular significance in this field are skyrmions, the most fundamental and extensively studied topological spin textures. First experimentally observed in magnetic systems in 2009 [10], skyrmions manifest as nontrivial spin textures in chiral magnets arising from the competition between Heisenberg exchange and Dzyaloshinskii-Moriya interaction (DMI) [11].

Recent studies have confirmed the emergence of skyrmions in 2D van der Waals (vdW) ferromagnets like $Fe_3GeTe_2$ [12-16], $Fe_5GeTe_2$ [17, 18] and $Cr_2Ge_2Te_6$ [19]. Notably, $Fe_{3-x}GeTe_2$ has revealed composite topological spin textures including skyrmion bags, skyrmioniums, and skyrmion sacks [5], demonstrating versatile functionalities for spintronic applications while paving new avenues in 2D topological magnetism research. Researchers have developed multiple control strategies — chemical doping [14], gate modulation [15], and pressure engineering [16]— enabling precise manipulation of magnetic configurations in 2D materials. These advances have significantly propelled the application of magnetic vdW materials in spintronics and magnetic storage technologies. However, the practical implementation faces critical challenges as most 2D magnetic systems exhibit Curie temperatures ($T_C$) below room temperature (RT), fundamentally limiting their viability for RT spin-based devices and information storage solutions.

A newly discovered vdW ferromagnet $Fe_{3-x}GaTe_2$ has recently been reported to exhibit a record-high $T_C$ of 350-380 K [20-41], establishing a novel RT operational platform for 2D magnetic material development and opening unprecedented opportunities for practical spintronic applications. Beyond its exceptional $T_C$ and pronounced perpendicular magnetic anisotropy, $Fe_{3-x}GaTe_2$ has attracted extensive research interest due to its diverse topological spin textures, including skyrmions [20-25], antiskyrmions [26], and composite skyrmions [27].

These topological spin textures not only contribute to the topological Hall effect in $Fe_{3-x}GaTe_2$ but also exhibit remarkable controllability [24-27]. Iron vacancies in $Fe_{3-x}GaTe_2$ critically stabilize topological spin textures by breaking inversion symmetry and generating DMI [20,24]. While vacancy concentration variations are hypothesized to govern distinct spin configurations, experimental validation of this correlation remains unexplored. Establishing this mechanistic link is crucial for tailoring topological evolution, particularly in high-order composite skyrmions, with significant implications for spintronic device engineering.

In this work, we report the experimental observation of multistage skyrmionic transitions in non-stoichiometric $Fe_{3-x}GaTe_2$ through combined magnetic force microscopy (MFM) and micromagnetic simulation, establishing a direct correlation between iron vacancy concentrations and the evolution of distinct topological spin textures. Firstly, the three typical magnetic states of stripe, striped skyrmionium, and striped skyrmion sack were distinctly observed in three iron-deficiency-specific regions of the same sample. Subsequently, the emergence and formation of these spin textures are further revealed by detailed zero-field-cooling (ZFC) and field-cooling (FC) measurements, indicating intrinsic topological-transition and trivial-transformation evolution processes. These observations provide a critical foundation for constructing an novel evolutionary pathway model. Finally, by selectively suppressing trivial transformation paths, the composite topological spin textures, including skyrmionium, skyrmion bag and sack states, have been controllably realized from the initial emergent skyrmion lattice state through an exclusive topological transition path. Our results not only reveal the tunable composite skyrmionic spin textures in $Fe_{3-x}GaTe_2$ at RT, but also highlight their great potential for applications in skyrmionic spintronic devices and proximity effects in 2D heterostructures.

# Result and discussion

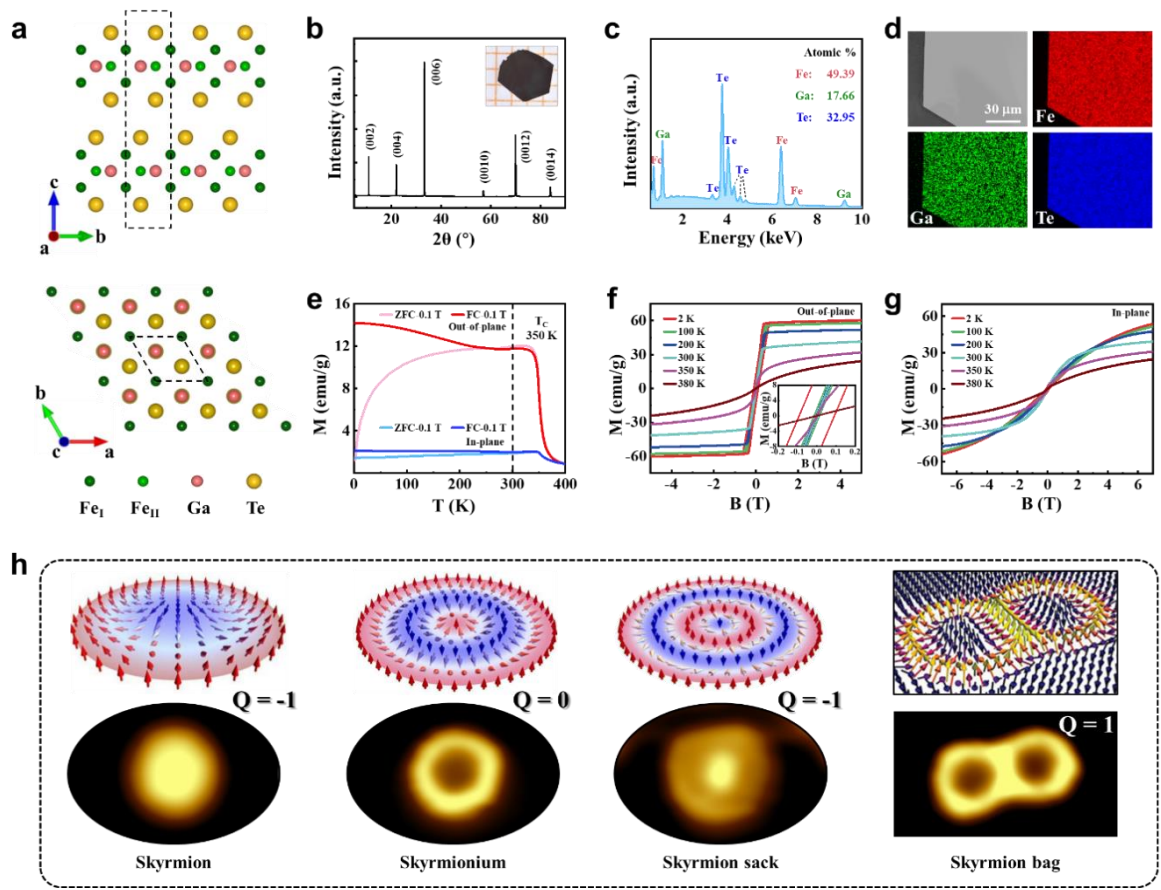

**Figure 1. Crystal characterization and magnetization measurement of $Fe_{3-x}GaTe_2$ single crystals. (a)** Side- and top-view structural models of $Fe_3GaTe_2$. **(b)** X-ray diffraction data of as-grown single crystal $Fe_{3-x}GaTe_2$ with an inset of its optical image. **(c,d)** EDS spectra (c) and elements mapping (d) of $Fe_{3-x}GaTe_2$. **(e)** Temperature-dependent magnetization curves measured with 0.1 T external field along *c*-axis and *ab*-plane for $Fe_{3-x}GaTe_2$. **(f,g)** Field-dependent magnetization curves at different temperatures with H//c (f) and H//ab (g) for $Fe_{3-x}GaTe_2$. **(h)** Schematic illustrations [5, 42] and MFM images of the obtained topological spin textures in $Fe_{3-x}GaTe_2$, including skyrmion (Q = -1), skyrmionium (Q = 0), skyrmion sack (Q = -1) and skyrmion bag (Q = 1).

As shown in Figure 1a, Fe$_3$GaTe$_2$ crystallizes in a hexagonal structure belonging to the *P6$_3$/mmc* space group, with lattice parameters a = b = 3.9860 Å, c = 16.2290 Å, α = β = 90°, γ = 120° [28]. The crystal architecture features a distinctive layered organization where Fe$_3$Ga slabs are sandwiched between two Te atomic layers, creating vdW gaps between adjacent Te layers. These structural units exhibit stacking alignment along the c-axis direction, possessing an interlayer spacing of approximately 0.78 nm [28]. To precisely regulate iron content, we developed an optimized growth protocol and successfully synthesized a series of Fe$_{3-x}$GaTe$_2$ single crystals with spatially tunable iron stoichiometry through chemical vapor transport method (see Methods). X-ray diffraction (XRD) analysis confirms the high crystallinity and phase purity of Fe$_{3-x}$GaTe$_2$ single crystals (Figure 1b). Energy Dispersive X-ray (EDX) elemental mapping revealed a homogeneous spatial distribution of Fe, Ga, and Te within specific regions, with quantitative ratios determined through spectral analysis (Figure 1c,d). This manuscript specifically selects single-crystal specimens exhibiting a distinct spatial Fe content distribution (2.92 - 2.96) for detailed investigation. This intrinsic inhomogeneity provides a unique platform to investigate the correlation between Fe content and topological spin textures within individual crystals.

Supplementary scanning tunneling microscopy investigations on non-stoichiometric Fe$_{3-x}$GaTe$_2$ have conclusively verified the presence of iron vacancy defects, with atomic-resolution imaging revealing characteristic lattice distortions consistent with missing Fe sites (Figure S1). The temperature-dependent magnetization (M-T) curves demonstrate a characteristic ferromagnetic (FM) phase transition in the Fe$_{3-x}$GaTe$_2$ single crystals, with a T$_C$ reaching approximately 350 K (Figure 1e), consistent with previously reported values [20, 24, 28]. Field-dependent magnetization (M-H) measurements across multiple temperatures reveal the c-axis orientation of the easy magnetization axis (Figure 1f,g), indicating pronounced perpendicular magnetic anisotropy. Notably, while weak FM hysteresis is observed at 2 K, this behavior progressively diminishes with increasing thermal excitation. The M-H curve measured above the T$_C$ demonstrates paramagnetic (PM) characteristics.

Prior to the detailed report, we establish a foundational framework by systematically classifying the distinct topological spin textures observed in Fe$_{3-x}$GaTe$_2$, which constitute the structural basis for the subsequent discussion of magnetic phenomena. Our magnetic characterization of $Fe_{3-x}GaTe_2$ reveals four distinct topological spin textures: (i) skyrmion, (ii) skyrmionium, (iii) skyrmion sack, and (iv) skyrmion bag, as shown in Figure 1h. Conventional skyrmions are stabilized by topological protection mechanisms, characterized by a quantized topological charge $Q = \pm 1$ (here we use -1 as an example). For skyrmionium, it is considered as a combination of two skyrmions with opposite topological charge, resulting in a total topological charge $Q = 0$ [2,43-47]. Skyrmioniums, being topologically trivial configurations, do not exhibit the characteristic skyrmion Hall effect associated with non-trivial topological charge transport. The spin configurations of skyrmion sacks and bags demonstrate enhanced structural complexity, representing topologically non-trivial high-order composite skyrmions. The controlled integration of multiple $Q = -1$ skyrmions into a skyrmionium induces the formation of skyrmion sack configurations, where the resultant topological charge $Q \leq -1$, depending on the number of skyrmions. Conversely, substituting these $Q = -1$ units with $Q = +1$ skyrmions generates skyrmion bag configurations, whose enhanced topological charge $Q \geq +1$. The formation of topological spin textures in $Fe_{3-x}GaTe_2$ arises from DMI [20,21,24], whereas stoichiometric $Fe_3GaTe_2$ exhibits centrosymmetric crystal symmetry that suppresses such chiral interactions. Consequently, increased non-stoichiometry (lower Fe content) will enhance the structural complexity of emergent spin textures through intensified symmetry breaking.

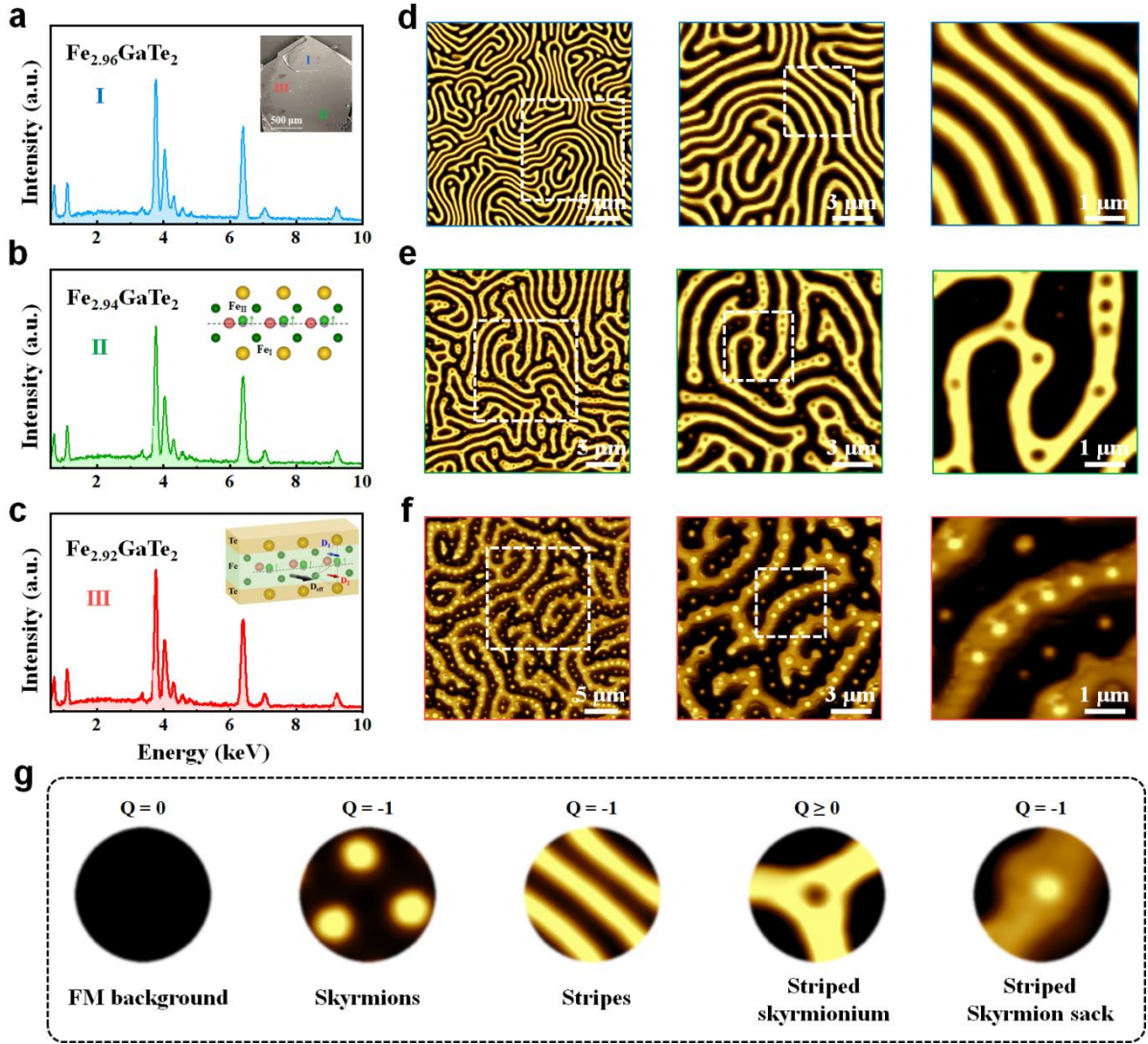

**Figure 2. MFM images of the region-specific spin textures in non-stoichiometric $Fe_{3-x}GaTe_2$ at RT. (a-c)** The three region-specific EDX data of the as-grown $Fe_{3-x}GaTe_2$ sample, marked on the inset SEM image in (a). The off-centered $Fe_{II}$ in the non-stoichiometric (Fe-deficient) $Fe_{3-x}GaTe_2$ contributes to the formation of non-centrosymmetric crystal structures (b), which directly leads to the introduction of effective DMI (c). **(d,e,f)** Multi-scale MFM images of regions I (d), II (e), and III (f) at RT. They exhibit spin textures of stripe domains, striped skyrmioniums, and striped skyrmion sacks, respectively. **(g)** The zoomed-in MFM images of the observed diverse spin textures.

To obtain distinct topological spin textures, we intentionally selected an $Fe_{3-x}GaTe_2$ single crystal with spatially inhomogeneous iron distribution. This strategic material selection allows systematic exploration of spin configurations while effectively eliminating most confounding factors. EDX analysis identified distinct chemical compositions of 2.96 (I), 2.94 (II), and 2.92 (III) for iron in three specific regions (Figure 2a-c). According to previous reports [20, 24], the presence of iron vacancies induces a displacement of $Fe_{II}$ from its central position, thereby reducing the symmetry of $Fe_{3-x}GaTe_2$ crystals and altering the space group to the non-centrosymmetric *P3m1* (Figure 2b inset and Figure S2). The off-centered $Fe_{II}$ atoms break the inversion symmetry, resulting in unequal **$D_1$** and **$D_2$** components that generate a non-zero effective DMI (**$D_{eff}$**) within each monolayer (Figure 2c inset and Figure S2). Therefore, depending on the iron vacancy concentration, the DMI in regions I to III was gradually enhanced.

Figures 2d-f present MFM images conducted at RT after ZFC on the three regions, revealing markedly different topological spin textures. For region I with the weakest DMI, the magnetic ground state manifests as a typical FM labyrinth domain, which is considered topologically trivial. The labyrinth domain observed in region I originates from the predominant roles of magnetocrystalline anisotropy and dipole-dipole interactions in governing domain evolution, which effectively quenches the weak DMI. As the DMI increases, region II exhibits the emergence of $Q = +1$ skyrmions within the striped domains, which we designate as striped skyrmioniums. In region III with the strongest DMI, striped skyrmioniums develops $Q = -1$ skyrmions within its interior, forming what we term striped skyrmion sacks. As a rarely reported unique spin texture, the formation of both striped skyrmionium and striped skyrmion sack is accompanied by changes in topological order, corresponding to a topological transition (Figure S3). It is important to note that the background signal intensity within the striped skyrmion sacks is greater than that outside, which can be attributed to a spurious elevation signal introduced by the MFM tip (see Figure S4 for details). These findings establish a direct correlation between iron vacancy concentration and topological spin textures in $Fe_{3-x}GaTe_2$. An increase in iron vacancies leads to a stronger DMI, resulting in a more complex spin texture of the magnetic ground state.

We extracted the distinct local spin textures emerging in the $Fe_{3-x}GaTe_2$ magnetic ground state, as shown in Figure 2g, which demonstrates the potential for tailoring composite skyrmionic spin textures. To gain a comprehensive understanding of the growth and evolution process of these topological spin textures, we conducted temperature-dependent MFM studies, using regions I and III as representatives.

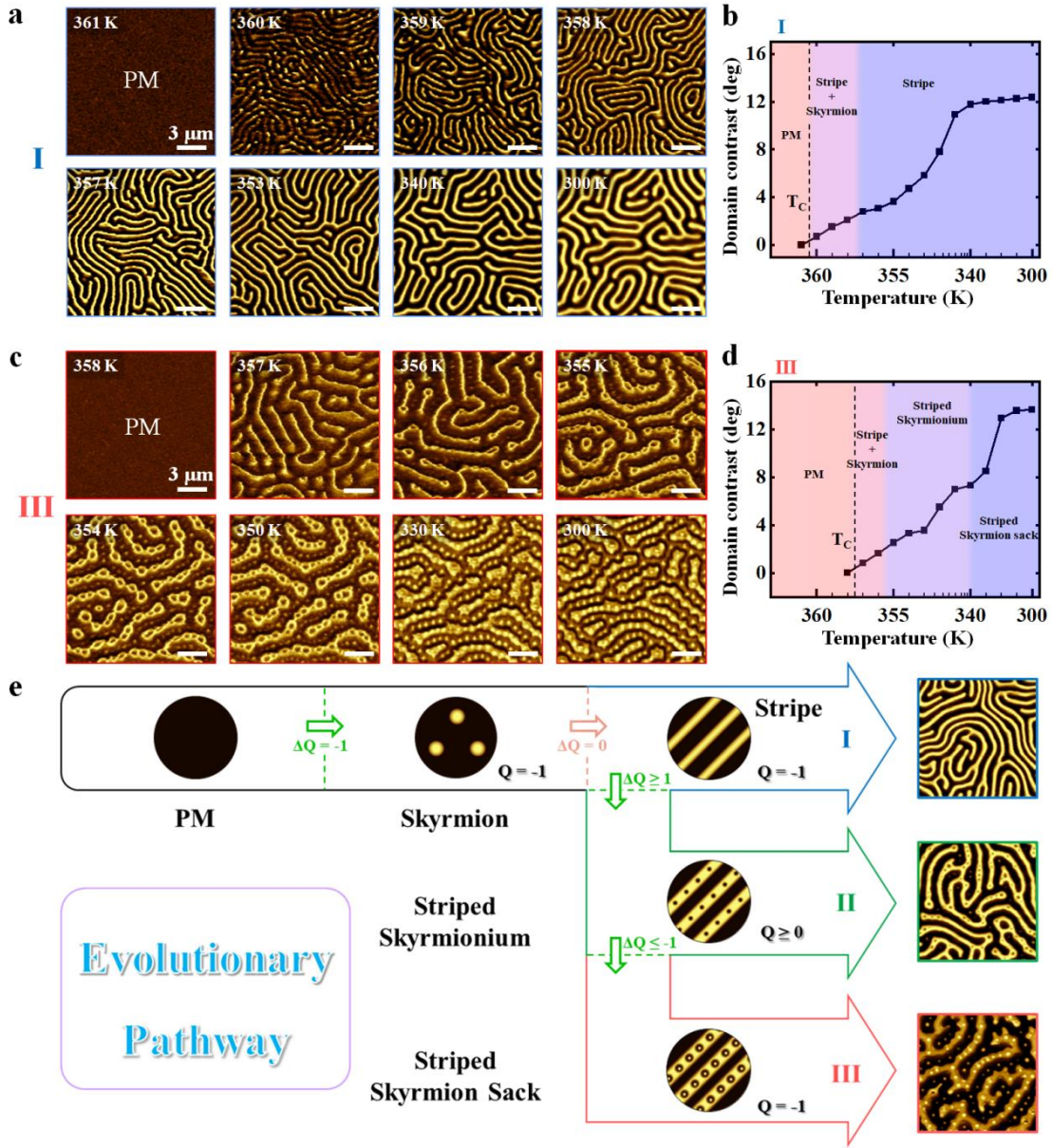

**Figure 3. MFM measurements of the temperature-dependent evolution of spin texture in $Fe_{3-x}GaTe_2$.** **(a-d)** Temperature-dependent MFM images and domain contrast curves of regions I (a,b) and III (c,d), showing distinct domain evolution pathways. **(e)** Schematic of the proposed stage-by-stage evolution of the observed complex topological spin textures in regions I, II, and III during the ZFC process. The green and pink arrows represent topological and trivial transitions, respectively.

Through precise temperature control (with 1 K accuracy), we performed continuous cooling MFM characterizations on regions I and III, with the experimental process schematically depicted in Figure S5. Region III (358 K) exhibits a modest decrease in $T_C$ compared to region I (361 K), correlating with its diminished iron content [24]. In region I, the magnetic domain evolution exhibits trivial behavior throughout the entire ZFC process (Figure 3a). At 361 K, region I exhibited a PM state with no discernible contrast in the MFM image. As the temperature decreased to 360 K, faint domain contrast began to emerge in the MFM image, revealing a mixed phase comprising both skyrmions and stripe domains. Upon further cooling, the domain contrast intensified gradually while the skyrmions progressively elongated and transitioned into stripe domains. This entire process, characterized as a trivial transition, occurred without any alteration in topological order. Ultimately, region I developed a well-defined labyrinthine stripe domain at RT. It is noteworthy that we propose the stripe domains evolved exclusively from skyrmion states, with a transitional pure skyrmion state potentially occurring within an extremely narrow thermal window preceding the skyrmion-stripe domain mixed phase. This transitional state, however, might have eluded experimental detection due to limitations in our temperature stabilization capability (< 1 K). The domain contrast curve throughout the cooling process is systematically presented in Figure 3b. Owing to the relatively straightforward evolution path of the magnetic domains, the domain contrast stabilizes at higher temperatures (approximately 340 K).

In contrast, the domain evolution in region III exhibits significantly greater complexity, manifesting as a series of topological transitions throughout the entire ZFC process (Figure 3c). Upon cooling just below the $T_C$ (357 K), region III reveals a mixed state comprising both skyrmions and stripe domains, analogous to region I. With further cooling, the DMI mediates the nucleation of skyrmions (Q = +1) within the stripe domains, forming a unique spin configuration of striped skyrmioniums. This structural progression corresponds to a topological transition accompanied by a variation in topological order. Ultimately, the striped skyrmioniums continue to generate skyrmions (Q = -1) to form the striped skyrmion sacks. The domain contrast curve throughout the cooling process is systematically presented in Figure 3d. The stabilization of domain contrast necessitates cooling to lower temperatures (320 K), owing to the increased complexity of the evolutionary pathway.

Based on the aforementioned experimental results, we established an evolutionary pathway model, as shown in Figure 3e. The magnetic domain evolution exhibits both trivial and topological transitions, with the variation in topological charge serving as the fundamental criterion for distinguishing between these two processes. Various topological spin textures exhibit distinct DMI thresholds, such that the corresponding topological transformations can only occur when the DMI surpasses these respective thresholds. At weak DMI intensities, which are insufficient to modify the topological charge of magnetic domains, only trivial transformations occur, leading to the formation of labyrinthine domains (region I). With increasing DMI strength, topological transformations emerge alongside trivial ones, giving rise to striped skyrmioniums (region II). As the DMI strength progressively increases, these striped skyrmioniums undergo further evolution into striped skyrmion sacks (region III). Therefore, we conclude that complex topological spin textures are not generated abruptly but rather evolve gradually through progressive mechanisms of topological charge evolution.

Building upon this model, the continued enhancement of DMI is anticipated to drive magnetic domains into more sophisticated spin textures through subsequent evolutionary processes. In other words, distinct topological spin textures can be controllably engineered by modulating the Fe content in the samples. Given that FC can effectively suppresses trivial transitions (eliminating stripe domains), our experimental results suggest that regions I to III are likely to exhibit extensive areas of well-defined composite skyrmionic spin textures under FC conditions. To this end, we subsequently conducted systematic FC characterizations in these three regions.

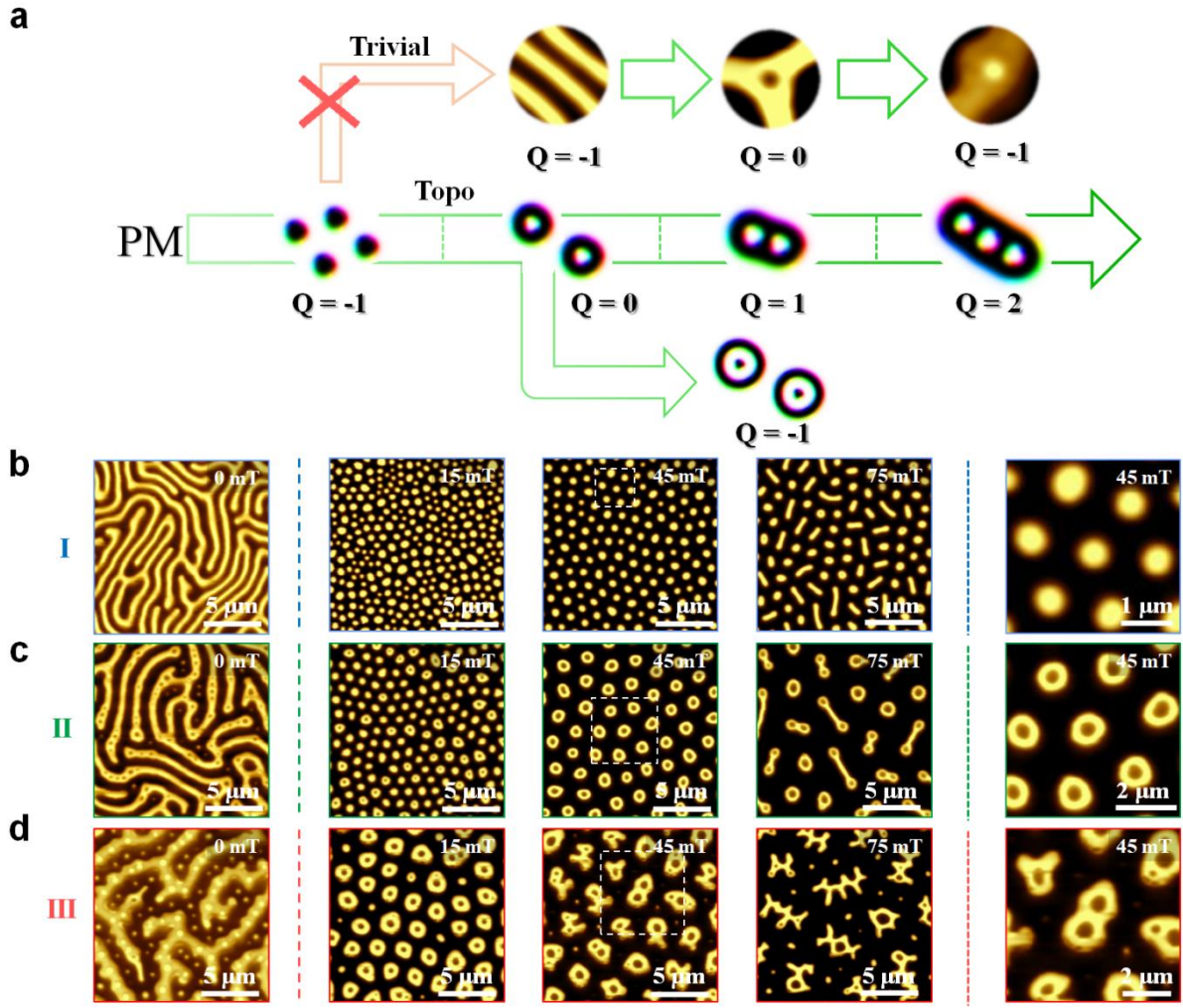

**Figure 4. MFM measurements of region-specific spin textures in Fe$_{3-x}$GaTe$_2$ after the FC process. (a)** Schematic of the topological (green arrow) and trivial (pink arrow) evolution paths for the distinct topological spin textures. **(b-d)** MFM images at RT of regions I, II, and III after ZFC (left) and FC (middle) processes, along with zoom-in images (right).

According to the evolutionary pathway model, trivial transitions are suppressed, leaving only topological transitions. The predicted evolution of topological spin textures under these conditions is shown in Figure 4a. The experimental protocol, as illustrated in Figure S6, involved cooling the system from high temperatures to eliminate memory effects. As anticipated, the experimental results demonstrate effective suppression of trivial transitions while clearly revealing the topological transition pathway, as shown in Figure 4b-d (see Figure S6 for additional data). In region I, after eliminating the trivial transformations, the complete skyrmion lattice remains preserved, thereby validating our previous hypothesis regarding the topological evolution pathway where the skyrmion lattice serves as the initial state of topological transformation. As the FC magnetic field increases, the skyrmion density decreases significantly, and the spin texture undergoes further evolution. The DMI constraints on topological transitions force a portion of skyrmions to undergo trivial transformations instead. This observation aligns well with our earlier discussions.

For region II, the enhanced DMI drives successive topological transitions. When the FC magnetic field increases from 15 mT to 45 mT, the magnetic configuration evolves from a coexistence state of skyrmions and skyrmioniums to a fully developed skyrmionium lattice. The emergence of such an extended skyrmionium lattice represents a remarkably rare phenomenon in 2D vdW systems. Subsequent experimental verification through temperature-dependent evolution studies definitively confirms that these large-scale skyrmionium lattices originate from the structural transformation of skyrmion lattices (Figure S7). As the FC magnetic field further increases, the topological transition and trivial transition occur simultaneously, with skyrmioniums transforming into skyrmion bags while exhibiting significant elongation. Region III demonstrates distinct behavior due to stronger DMI, which facilitates easier topological transitions. Remarkably, even under a relatively weak FC magnetic field (15 mT), the system enters skyrmionium spin textures. The presence of external magnetic fields notably suppresses skyrmion sack formation. Upon field removal, the skyrmionium lattice in region III undergoes further evolution into a skyrmion sack lattice, while region II maintains structural stability (Figure S8) - a clear demonstration of DMI's critical role in governing topological evolution pathways. As FC magnetic fields intensify in region III, high-order topological transitions begin to emerge, characterized by high-order skyrmion bags and increasingly complex magnetic domain structures. The dynamic processes and types of these high-order topological transitions exhibit significantly greater complexity, which we will briefly discuss in the following section.

This progressive evolution of FC highlights the delicate balance between DMI strength and magnetic field effects in shaping topological spin textures. Thus far, we have established a comprehensive evolutionary pathway model. External parameters, including DMI intensity, temperature, and magnetic fields, demonstrate crucial regulatory effects on pathway selection. Specifically, DMI strength determines the terminal state of evolution, while temperature modulates the progression stage within a given pathway. The applied magnetic field not only suppresses trivial transitions but also synergistically modulates the evolution dynamics in conjunction with DMI. This established framework enables precise control over evolutionary trajectories, thereby facilitating the tailored creation of composite skyrmionic spin textures.

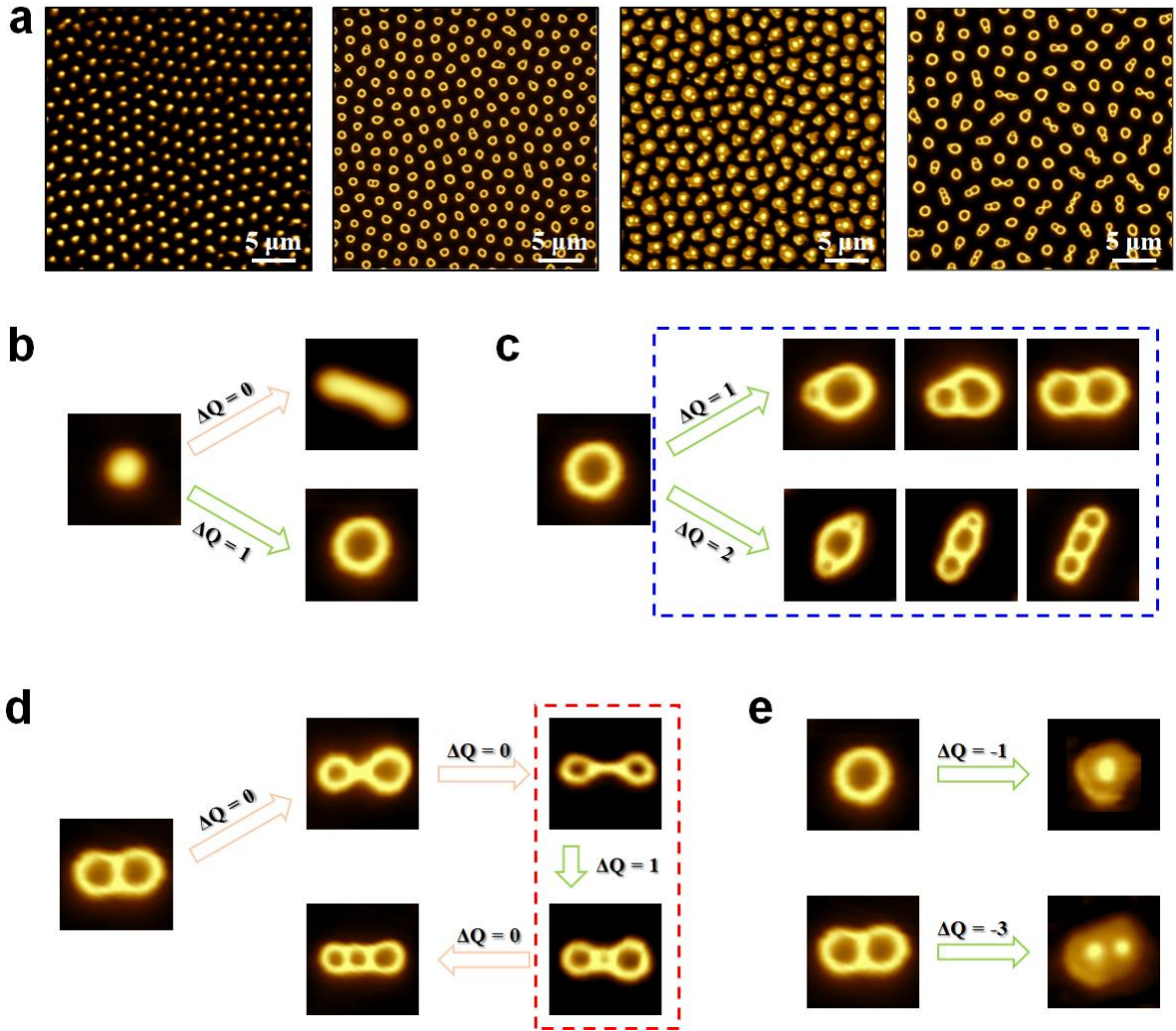

**Figure 5. Tailoring the composite skyrmionic spin textures in Fe$_{3-x}$GaTe$_2$. (a)** MFM images of the tailored large-area uniform skyrmion lattice, skyrmionium lattice, skyrmion sack lattice, and skyrmion bag lattice. **(b-e)** MFM images of a series of spin textures, showing the trivial (pink arrow) and topological (green arrow) transition paths of a single skyrmion. The blue and red dashed boxes represent two distinct high-order topological transitions, defined as the chiral kink topological transition (c) and the trivial growth topological transition (d), respectively.

Leveraging the evolutionary pathway model, various large-area, stable composite skyrmion lattices have been successfully tailored at RT, as demonstrated in Figure 5a (Figure S9). Among them, the single-topological-charge ($|Q| \leq 1$) skyrmion lattice exhibits higher uniformity. Starting with a skyrmion lattice ($Q = -1$), topological transformation can lead to the formation of a skyrmionium lattice ($Q = 0$) and a single skyrmion sack lattice ($Q = -1$). The global nature of single-topological-charge transformations plays a key role in facilitating the formation of uniform lattices. However, higher-order topological transitions (skyrmion bag lattice, $Q \geq 1$), due to their increased complexity, exhibit reduced uniformity.

Taking the single skyrmion as a representative case, the evolutionary pathway model encapsulating high-order topological transitions is schematically presented in Figure 5b-e (Figure S10). The high-order topological transitions from a skyrmionium to a skyrmion bag can be classified into two distinct categories: chiral kink topological transitions (blue dashed box) and trivial growth topological transitions (red dashed box). For chiral kink topological transitions, as the nomenclature suggests, chiral kinks initially develop in the skyrmionium configuration, followed by the progressive expansion of these kinks, leading to bag formation [48]. This transition mechanism preferentially generates chain-like spin textures. For another type of trivial growth topological transition, a trivial transition occurs first, leading to the emergence of stripe domains, followed by the generation of new skyrmions within the stripe domain regions, thereby forming high-order skyrmion bags. The dynamics of high-order topological transitions are currently under investigation, with more detailed explorations to be conducted in future studies. It is noteworthy that the occurrence of high-order topological transitions exhibits enhanced stochasticity, manifesting as statistically independent events for individual skyrmions, which consequently leads to the non-uniformity of the skyrmion bag lattice. Based on the above results, we have systematically analyzed and summarized the relationship between trivial and topological transitions across distinct topological spin textures, as detailed in Figure S11.

The in-situ field-dependent characterization also provides additional support for the evolutionary pathway model, as shown in Figure S12, where the field-increasing process of the skyrmion sack lattice aligns with our inferred evolutionary trajectory. Additionally, the $Fe_{3-x}GaTe_2$ thin film was characterized, as shown in Figure S13. The enhanced shape anisotropy in these thin-film systems imposes a stronger critical threshold for DMI strength - only when surpassing this confinement effect can the magnetic ground state support stabilized, more complex topological spin textures, which is in agreement with the findings from micromagnetic simulations (Figure S14). The strategic enhancement of DMI through Fe defect engineering is highly likely to induce domain evolution from conventional skyrmion lattices to high-order composite configurations in thin-film systems. Through the evolutionary pathway model, we have successfully engineered tailored composite skyrmionic spin textures. These topological spin textures not only demonstrate significant potential for advancing topological quantum computing (Figure S15)— particularly through the construction of skyrmion-vortex pairs potentially hosting Majorana zero modes [49-51]— but may also emerge as novel conceptual frameworks for next-generation information storage or reservoir computing devices.

**Conclusions**

In summary, we conducted systematic MFM investigations on non-stoichiometric $Fe_{3-x}GaTe_2$ exhibiting spatially heterogeneous iron distribution. The variation in DMI strengths across distinct crystalline domains directly governs the formation of region-specific topological spin textures, a finding that is consistent with the results of micromagnetic simulations. Detailed ZFC and FC characterizations conducted in various regions have facilitated the successful establishment of a model for the evolutionary pathway of spin textures. Leveraging this model, we successfully engineered tailored composite skyrmion lattices with programmable topological charges. These findings not only provide crucial experimental references for fabricating high-order topological spin textures but also open new avenues for developing skyrmion-based spintronic devices.

**Acknowledgments**

This project was supported by the National Key R&D Program of China (MOST) (Grant No. 2023YFA1406500), the National Natural Science Foundation of China (NSFC) (No. 21622304, 61674045, 11604063, 11974422, 12104504, 12374199), the Strategic Priority Research Program (Chinese Academy of Sciences, CAS) (No. XDB30000000), and the Fundamental Research Funds for the Central Universities and the Research Funds of Renmin University of China (No. 21XNLG27). Y. Y. Geng was supported by the Outstanding Innovative Talents Cultivation Funded Programs 2023 of Renmin University of China. This paper is an outcome of "Study of Exotic Fractional Magnetization Plateau Phase Transitions and States in Low-dimensional Frustrated Quantum Systems" (RUC24QSDL039), funded by the "Qiushi Academic-Dongliang" Talent Cultivation Project at Renmin University of China in 2024.

## Materials and Methods

### Single preparation and characterization

The growth of $Fe_{3-x}GaTe_2$ single crystals was accomplished through chemical vapor transport (CVT) technique. The growth process initiated with the prepared mixture comprising Fe powder (99.95%), Ga pellets (99.9999%), and Te lumps (99.999%) in a stoichiometric molar ratio of 3:1:2 within an argon-filled glovebox. Subsequently, the prepared mixture, along with a precise quantity of the transport agent $NH_4Cl$, was hermetically sealed within an evacuated quartz tube ($<10^{-4}$ Pa). The sealed quartz tube was then positioned within a two-zone tubular furnace. The temperatures were incrementally raised to 880/780 ºC at a rate of 2 ºC /min, then sustained at the temperature for a duration of two weeks, followed by gradual cooling to ambient temperature. Ultimately, a hexagonal $Fe_{3-x}GaTe_2$ single crystal with metallic luster was attained. The phase purity of the single crystal was analyzed by the powder X-ray diffraction with the Cu Kα radiation. The stoichiometric ratio was determined by the scanning electron spectroscopy (SEM) equipped with EDS. The magnetization of the single crystal was measured by a SQUID magnetometer (Quantum Design MPMS-3).

### MFM measurements.

All MFM characterizations were performed at both RT and elevated temperatures using two independent systems. The majority of MFM measurements were performed using a commercial AFM system (Park NX10, Park Systems) integrated with a precision heating stage, in conjunction with commercial AFM probes (Nanosensors, PPP-MFMR). The MFM images were captured in the dual pass MFM mode with lift height ~50 nm, in which the first scanning process is performed at the resonance frequency of the magnetic tip (approximately 65 kHz). In this system, the method of applying a magnetic field was placing permanent magnet pieces (approximately 150 Gs each) underneath the sample. A high-precision Gauss meter is employed to accurately determine the magnetic field in which the sample is situated. High-precision increasing-magnetic-field MFM measurements were performed using a commercial magnetic force microscope (attoAFM I, Attocube) employing a commercial magnetic tip (Nanosensors, PPP-MFMR) based on a closed-cycle He cryostat (attoDRY2100, Attocube). The MFM images were captured in a constant height mode with the scanning plane nominally ~100 nm above the sample surface. The MFM signal, i.e., the change in the cantilever phase, was proportional to the out-of-plane stray field gradient. The dark (bright) regions in the MFM images represented attractive (repulsive) magnetization, where the magnetization was parallel (antiparallel) to the magnetic tip moments.

### Micromagnetic simulations.

Micromagnetic simulations based on the Landau-Lifshitz-Gilbert (LLG) equation are performed using the Mumax3 software [52]. Given the variability in magnetic states across different regions of the sample, it is necessary to consider distinct parameters for each specific area. To construct as comprehensive a magnetic ground state phase diagram as possible, we maintain a constant exchange interaction for the material while varying the DMI and magnetic anisotropy. The exchange interaction strength was set as A = 5.0 pJ/m, which is a parameter based on bulk $Fe_3GaTe_2$ crystal at RT. The DMI varies within the range of 0.2 - 1 $mJ/m^2$ and the uniaxial anisotropy constant Ku varies within the range of 0.4 - $1.6 \times 10^5$ J/m$^3$. The Gilbert damping constant is set as $\alpha = 0.5$, and the simulation cell is set to $5 \times 5 \times 5$ nm$^3$. The in-plane dimensions of the simulations are set to $2000 \times 2000$ nm$^2$.